\documentstyle[12pt,fleqn]{article}
\input epsf
\def\singlespace {\smallskipamount=3.75pt plus1pt minus1pt
                  \medskipamount=7.5pt plus2pt minus2pt
                  \bigskipamount=15pt plus4pt minus4pt
                  \normalbaselineskip=15pt plus0pt minus0pt
                  \normallineskip=1pt
                  \normallineskiplimit=0pt
                  \jot=3.75pt 
                  {\def\smallskip {\vskip\smallskipamount}}
                  {\def\medskip   {\vskip\medskipamount}}
                  {\def\bigskip   {\vskip\bigskipamount}}
                  {\setbox\strutbox=\hbox{\vrule 
                    height10.5pt depth4.5pt width 0pt}}
                  \parskip 7.5pt
                  \normalbaselines}

\def\doublespace {\smallskipamount=7.5pt plus2pt minus2pt
                  \medskipamount=15pt plus4pt minus4pt
                  \bigskipamount=30pt plus8pt minus8pt
                  \normalbaselineskip=30pt plus0pt minus0pt
                  \normallineskip=2pt
                  \normallineskiplimit=0pt
                  \jot=7.5pt
                  {\def\smallskip {\vskip\smallskipamount}}
                  {\def\medskip   {\vskip\medskipamount}}
                  {\def\bigskip   {\vskip\bigskipamount}}
                  {\setbox\strutbox=\hbox{\vrule 
                    height21.0pt depth9.0pt width 0pt}}
                  \parskip 15.0pt
                  \normalbaselines}

\def\be{\begin{equation}}
\def\ee{\end{equation}}
\def\bea{\begin{eqnarray}}
\def\eea{\end{eqnarray}}

\def\sect #1{\setcounter{equation}{0}}

\begin{document}
\singlespace
\begin{center}
{\Large{Physical nature of the central singularity in spherical collapse}}
\end{center}
\vspace{1.0in}
\vspace{12pt}

{\begin{center}
{\large{ S. S. Deshingkar\footnote{E-mail : shrir@tifr.res.in}, 
P. S. Joshi\footnote{E-mail : psj@tifr.res.in}, and I. H. Dwivedi
\footnote{E-mail : dwivedi@tifr.res.in}\\
Theoretical Astrophysics Group\\
Tata Institute of Fundamental Research\\
Homi Bhabha Road, Colaba, Bombay 400005, India.}}
\end{center}

\vspace{1.3in}

\begin{abstract}
\doublespace
We examine here the nature of the central singularity forming in the 
spherically symmetric collapse of a dust cloud and it is shown that this is 
always a strong curvature singularity where gravitational tidal forces 
diverge powerfully. An important consequence is that the nature of the naked 
singularity forming in the dust collapse turns out to be stable against the 
perturbations in the initial data from which the collapse commences.

\end{abstract}
\newpage

The final outcome of the gravitational collapse of a massive
matter cloud is an issue of great interest from the perspective of 
gravitation theory as well as its possible astrophysical implications.   
When there is a continual collapse without any final equilibrium for the 
cloud, a black hole may form where the superdense regions of the matter 
are hidden away from the outside observer within an event horizon of 
gravity or, depending on the nature of the initial data and the possible 
evolutions, a naked singularity could result as the end product of such
a collapse.

In this context, spherical dust collapse in general relativity has been 
studied quite extensively. The prescription of matter as pressureless dust 
could perhaps be regarded as somewhat idealized. However, some authors 
have considered dust as a good approximation of the form of matter in the
final stages of collapse$^1$. All the same, the study of
dust models over the decades has led not only to important new advances
and insights into various aspects of gravitational collapse theory, but 
has also laid
the foundation of black hole physics. The class of solutions representing an 
inhomogeneous spherically symmetric dust cloud was discovered and studied 
by Tolman, Bondi, and Lema{\^i}tre$^2$. In addition to its wide use in 
cosmology to model the universes which admit inhomogeneities, these 
models have been used extensively to investigate gravitational collapse
in general relativity. For example, the special case of homogeneous dust ball 
was studied by Oppenheimer and Snyder$^3$, which led to the concept and 
theory of black holes. This model also inspired the  cosmic censorship 
conjecture of Penrose$^4$, which basically states that singularities which 
are visible to outside observers, either locally or globally, cannot develop 
in gravitational collapse from regular initial data in a stable manner.

In the case of formation of a naked singularity in gravitational collapse, 
two aspects which become most significant are the strength of such a 
singularity in terms of the behavior of the gravitational tidal forces in 
its vicinity and its stability properties. Numerous papers on the strength
have led to the classification of various subclasses within the
Tolman-Bondi-Lema{\^i}tre (TBL) spacetimes, where the central singularity 
is either shown to be
weak or strong. Thus there is considerable debate on the genuineness and
physicality of the central singularity in dust collapse, 
especially for the cases where existing literature
points towards the singularity being gravitationally weak (see, e.g.,
Refs [5-9], and references therein).  
The importance of this lies in the fact that even if a naked singularity 
occurs rather than a black hole, if it is gravitationally weak in some 
suitable sense, it may not have any physical implications and it may 
perhaps be removable by extending the spacetime through the same$^{10}$. 
Such, for example, would be the case for the shell-crossing 
naked singularities $^{11}$, 
caused by the crossings of dust shells, through which a continuous extension 
of the metric has been shown to exist in certain cases, and the field 
equations hold in a distributional sense. Similarly, if a naked singularity 
is not stable in some well-defined sense, it may not have any significant 
physical consequences. The purpose of this paper 
is to investigate these 
aspects related to the central naked singularity forming in the dust 
collapse. As opposed to the shellcrossings mentioned above, the physical 
area of the dust shells vanish in this case when the matter shells 
collapse into the singularity at the center. We show here that whenever such 
a singularity forms as the end product of the collapse, it is always 
gravitationally strong. Certain stability aspects related to naked 
singularities are then pointed out.

The gravitational strength of a spacetime singularity is conveniently 
characterized in terms of the behavior of the linearly independent Jacobi
fields along the timelike or null geodesics which terminate at the 
singularity in the past or future$^{12}$. In particular, a causal
geodesic $\gamma(s)$, incomplete at the affine parameter value $s=s_0$, 
is said to terminate in a {\it strong curvature singularity} at $s=s_0$
if the volume three-form $V(s)= Z_1(s)\wedge Z_2(s)\wedge Z_3(s)$ (in the case
of a null geodesic, this will be defined as a two-form) vanishes in the limit
as $s\to s_0$ for {\it all} linearly independent vorticity free Jacobi fields 
$Z_1, Z_2, Z_3$ along $\gamma(s)$. A sufficient condition for the causal
geodesic $\gamma(s)$ to terminate in a strong curvature singularity$^{13}$ is 
that in the limit of approach to the singularity, we must have along $\gamma$,
\be
(s_0-s)^2 R_{ab}V^aV^b=const>0,
\ee
where $V^a$ is the tangent vector to the 
geodesic. This expresses the requirement on the strength in terms of the 
rate of divergence of the Ricci tensor along the given trajectory.
Essentially, the idea captured here is that in the limit of approach to
such a singularity, the physical objects get crushed to a zero size,
and so the idea of extension of spacetime through it would not make sense,
characterizing this to be a genuine spacetime singularity (we refer to 
Ref. [12] for a further discussion).

We now discuss the structure and strength 
of the singularity forming in the TBL collapse, 
an issue which is all the more important when it is naked. These models are 
fully characterized by the initial data, which corresponds to two 
free functions, namely, the initial density function $\rho(r)$ [or equivalently
the mass function $F(r)$] and the velocity or energy function $f(r)$, 
specified at the initial spacelike surface $t=t_i$ at the onset of collapse. 
Consider the dust collapse described by the TBL metric,
\be
ds^2= -dt^2 +{{R'^2}\over{1+f}}dr^2 + R^2d\Omega^2.
\ee
The energy-momentum tensor is that of dust
\be
T^{ij}=\epsilon\delta^{i}_{t} \delta^{j}_{t},
\epsilon=\epsilon(t,r)={F'\over R^2 R'},
\ee
where $\epsilon$ is the energy density, and $R =R(t,r)$ is given
by
\be
\dot R^2 = f(r) + {F(r)\over R}.
\ee
Here the dot and the prime denote partial derivatives with respect to the 
parameters $t$ and $r$, respectively, and for collapse we have $\dot R <0$. 
Using the remaining coordinate freedom left in the choice of rescaling
of the coordinate $r$ we set at the initial time $t=0$,
\be
R(0,r)=r.
\ee
The functions $F(r)$ and $f(r)$ are the mass and the energy functions 
respectively, referred to above. We have 
\be
\rho(r)\equiv\epsilon(0,r)={F'\over r^2}\Rightarrow F(r)=\int{\rho(r)r^2dr}.
\ee
Since at the onset of collapse at the initial surface $t=0$ the 
spacetime should be singularity free, it follows that $F(r)=r^3h(r)$ with
$F(0)=0$, 
and that the initial density $\rho(r)$ must at least be a $C^1$ function 
including the center at $r=0$, i.e. 
\be
\rho(r)=\rho_c{\big [} 1-\rho_1 r^n g(r){\big ]},
\ee
where $g(r)$ is at least $C^0$ function with $g(0)=1$, $n\ge 1$, and 
$\rho_1 >0$ are constants. The term $r^n$ is the first nonvanishing higher
order term in density at $r=0$. Thus, if $\rho'(0) \ne 0$, then $n=1$.
For physical reasons we take that $\rho(0)=\rho_c\ne 0$. 
The shell-focusing singularity
appears at $R=0$ along the curve $t=t_s(r)$ such that $R[t_s(r),r]=0$.
The central singularity at $r=0$ appears (for the case $f=0$) at the time
\be
t_s=t_s(0)={2\over\sqrt{3\rho_c}}. 
\ee
It is known$^{5,8}$ that given any initial density profile for such 
a cloud, one can always choose a corresponding velocity function (or vice
versa) describing the infall of the matter shells,
such that either a black hole or a central naked singularity would develop 
as desired as the outcome of the collapse. Thus a naked singularity 
would occur in a wide range of TBL spacetimes depending on the choice of
the initial data. The structure of this naked singularity for the 
marginally bound, time symmetric case, 
when the density was taken to be an even smooth 
function [i.e., $\rho'(0)=0$ and all the other odd derivatives of density 
also vanish] was 
analyzed analytically by Christodoulu$^{14}$, and later generalized to 
a wider class of TBL spacetimes with the same assumption of even 
smooth functions by Newman$^{6}$, who also showed the singularity 
to be gravitationally weak along the radial null geodesics terminating 
at the central singularity. The formation and structure of this 
naked singularity was then analyzed$^{8}$ 
with the only assumption of $C^2$ differentiability of the initial mass and 
velocity functions. Several later papers [see, e.g., Ref. [9], and references
therein) analyzed the precise 
role of the first three derivatives at the center of the initial 
density function towards determining 
the nature and structure of this naked singularity. 
The conclusion prevailing presently from all of this analysis is that 
in cases when the first two derivatives of the density at the center were 
nonvanishing, the naked singularity was gravitationally weak, because 
the curvature strength along the outgoing radial null geodesics did not 
diverge sufficiently fast. In other words, the central singularity formed 
in a dust collapse is naked but gravitationally weak in cases (i) $\rho'(0)<0$, 
(ii) $\rho'(0)=0,\,\rho''(0)<0$, however, gravitationally strong and naked 
in the cases when (iii) $\rho'(0)=\rho''(0)=0$, with $\rho'''(0)<0$ and 
less than a certain maximum.

Physically it is argued sometimes that the initial density 
distribution must be an even smooth function of $r$, therefore for all 
such cases the singularity would be weak. Furthermore, from the 
point of view of stability, any slight perturbation of a strong 
curvature singularity would involve the introduction 
of the term where $\rho'(0)\ne 0, \rho''(0)\ne 0$, and the formation of 
a strong curvature naked singularity would not be stable in this sense. 
As emphasized earlier, one could possibly remove a weak naked singularity
by possibly extending the spacetime, which, however, is not possible for 
a strong curvature singularity. Though it has been shown$^9$ that 
for any given arbitrary initial density distribution one could always 
choose the velocity function to be such that the resulting singularity 
is a strong curvature naked singularity, it is possible that 
the measure of such profiles may be vanishingly small 
in the space of all initial data, and in this sense such a 
strong naked singularity
may not be stable. Thus it could be argued that from the point of view of both 
the stability and curvature strength the naked singularities forming in 
the dust collapse may have very limited bearing on cosmic 
censorship hypothesis.

We shall show, however, that the structure of the central 
shell-focusing singularity is rather more complex than thought earlier.
Considering the marginally bound case ($f=0$) for the sake of clarity,
even the case of radial null geodesics can be seen to be quite involved.
The geodesic equation in this case, using Eq.(2) and an integration of Eq.(4),
is written as
\be
{dt\over dr}= {\sqrt{rF} -{1\over2}F't \over \sqrt{F}[r^{3/2}- {3\over2}
\sqrt{F}t]^{1/3}}.
\ee 
It is seen that in the limit of approach to the singularity both the
numerator and denominator vanish in the above equation, and therefore 
even in the simplest case of radial null geodesics the nature of the 
singularity turns out to be rather complex, which is a node for 
the above first order equation,
giving rise to a complicated topology of integral curves near the singularity. 
All the analysis and related conclusions so far on this naked singularity 
have been based on trying to understand only some of the 
families of these null geodesics.
Because of the complex and difficult nature of the geodesic equations,
especially in the timelike case, the discussion on them has been avoided so
far in the literature. It is, therefore, proposed to investigate here 
the nature of timelike geodesics
terminating in the naked central singularity. We find that this 
changes the existing perception on the nature of the central naked singularity
as discussed above, and in fact we get timelike radial geodesics 
along which the curvature growth is powerful enough near the singularity
to satisfy the strong curvature condition above.

Let us consider the timelike causal curves in the
TBL spacetime. Let $U^a=(dx^a/ds)$ ($s$ being the proper time along the 
trajectory) be the tangent to a timelike geodesic, 
satisfying $U^aU_a=-1$. We can express radial timelike geodesics as
\be {dU^t\over dr}\pm \dot R' \sqrt{{(U^t)^2-1\over 1+f}}=0,  \ee
\be {dt\over dr}={U^t\over U^r}=\pm{R'\over \sqrt{1-{1\over(U^t)^2}}
\sqrt{1+f}}
\Rightarrow {dR\over dr}=R'{\bigg(}1\mp{\sqrt{f+{F\over R}}\over 
\sqrt{1-{1\over (U^t)^2 }}\sqrt{1+f}}{\bigg)},\ee
where $\pm$ represent outgoing and ingoing curves and we have used
the variable $R$ instead of $t$ in the last equation. The solutions
to the set of two 
differential equations above in the form of $R(t,r)=f_1(r)$ and 
$U^t=U^t(R,r)$, describe the trajectories of particles 
following  timelike geodesics.

Because of the nonlinearity, solutions to the above differential 
equations are not available in general. However, there is an exact 
solution of the above described by
\be
U^a={dx^a\over ds}=\delta^a_t,\quad r=0.
\ee
This is an ingoing radial timelike geodesic $\gamma(s)$ which
in $(t,r)$ plane is described by $r=0$, terminating at the central 
singularity at the coordinate time $t=t_s(0)$,
corresponding to a particle at the center $r=0$ following a timelike 
geodesic and terminating in the singularity in future. 
The tangent to this timelike geodesic is given by the above equation,
where $s$ is the proper time. The
equation of the trajectory is simple and is given by
\be t_s-t=s_0-s,\quad r=0 \ee
where $s_0$ is the proper time when the particle crashes into the
central singularity at $r=0$ at the coordinate time $t_s(0)$.

It is now possible to show that the above is not the only geodesic 
terminating in the future into the singularity. In fact, there are 
infinitely many families of future 
directed null and timelike geodesics from the past of the central singularity 
at $r=0,t=t_s$, terminating at the singularity in future. To see this,
consider the past $I^{-}(\gamma)$ of the timelike geodesic $\gamma(s)$
given above. Since $\gamma(s)$ is future endless and future incomplete,
terminating in the central singularity at $r=0$, the set $I^{-}(\gamma)$ 
is a terminal indecomposable past (TIP) set as characterized by Geroch, 
Khronheimer and Penrose$^{15}$. The boundary of this past set is then
generated by null geodesics which are endless in future, terminating at    
the central singularity, by the theorem (2.3) of Ref. [15].
It follows that infinitely many families of null geodesics terminate at 
the singularity in future. If the collapse develops from the initial
spacelike surface $t=0$, all these null geodesics meet this surface in the
past in the TBL models. Consider now any event $p$ in  $I^{-}(\gamma)$.
This later set being a TIP, the past of $p$, $I^{-}(p)$ must be contained 
in  $I^{-}(\gamma)$. Thus all the past directed nonspacelike curves from
$p$ must meet the initial surface $t=0$, and so the set  $I^{-}(\gamma)$ 
becomes a subset of the domain of dependence (see, e.g., Ref. [16] for 
definitions 
and further details) of the initial surface $t=0$ in the spacetime. 
Then, given any event $p$ in  
$I^{-}(\gamma)$ and any other event $q$ in the future of $p$ in the same 
set, there exists a nonspacelike geodesic from $p$ to $q$. Choosing now 
$q$ to be closer and closer to the singularity, and in the limit 
approaching the same, we see that there are nonspacelike geodesics from 
$p$ entering every neighbourhood of the central singularity, and thus 
terminating in future only at the singularity. Thus, there will be 
infinitely many future directed nonspacelike geodesics, both null and 
timelike, which terminate at the central singularity in future.

One can also analyze equations (10),(11) further to understand better the 
structure of timelike geodesics terminating at the central singularity in 
future. Another such solution describing ingoing timelike causal 
particles is given, in the neighbourhood of the central singularity
(for $n<3$), as 
\be R(t,r)=X_0r^{1+2n/3},\quad U^t=1+{1\over 8}
{\bigg (} {b\over 1-n/3} {\bigg )}^2  r^{2(1-{n\over 3})}
+O(r^m), \ee
where
$$
{X_0}^{3/2}=\cases { {\rho_1\over 4}, & $n=1$, \cr
		-{n\rho_1\over (3+n)(n-1)}
			 (2\mp\sqrt{2(3-n)}), & $n\ne 1$, \cr}
$$
$$
b={\bigg(}2- {n\rho_1\over (3+n){X_0}^{3/2}}{\bigg)}
\sqrt{\rho_c\over 3X_0}.
$$
The trajectories described by the above equation actually represent
a family in the spacetime of infinitely many ingoing radial timelike geodesics 
($\theta=const$, $\phi=const$), terminating at the central singularity 
in future. In fact, the above timelike geodesics form a part of a 
larger family of timelike geodesic curves in the $(t,r)$ plane that 
terminate in the singularity and are given in the near regions of the 
singularity by
\be U^t={1+C^2\over 2C} \pm r^{1-n/3}{1-C^2\over 2C} 
\sqrt{\rho_c\over 3X_0} + O(r^m), \ee
\be R(t,r)=X_0r^{1+2n/3}, \ee

where $X_0^{3/2}={3\rho_1 \over 2(3+n)}$, $C\ne 1$ is constant, and 
$\pm$ represent 
outgoing and ingoing geodesics.
Figure 1 describes a typical radial timelike geodesic terminating at the 
central singularity. For the case $n\ge 3$ the singularity is already 
known to be strong$^{5,9}$, and families of null and timelike geodesics 
terminate at the singularity. 
We have discussed the  marginally
bound (i.e., $f=0$) case above for the sake of simplicity and clarity, however,
for the nonmarginally bound case as well one gets similar results. 
As pointed out earlier, the solutions describing the particle trajectories in
the $(t,r)$ plane terminating in the singularity are not limited just to the
solutions given by Eqs.(12)-(16), but there will be many more such families. 

\begin{figure}[h]
\parbox[b]{7.99cm}
{
\epsfxsize=7.95cm
\epsfbox{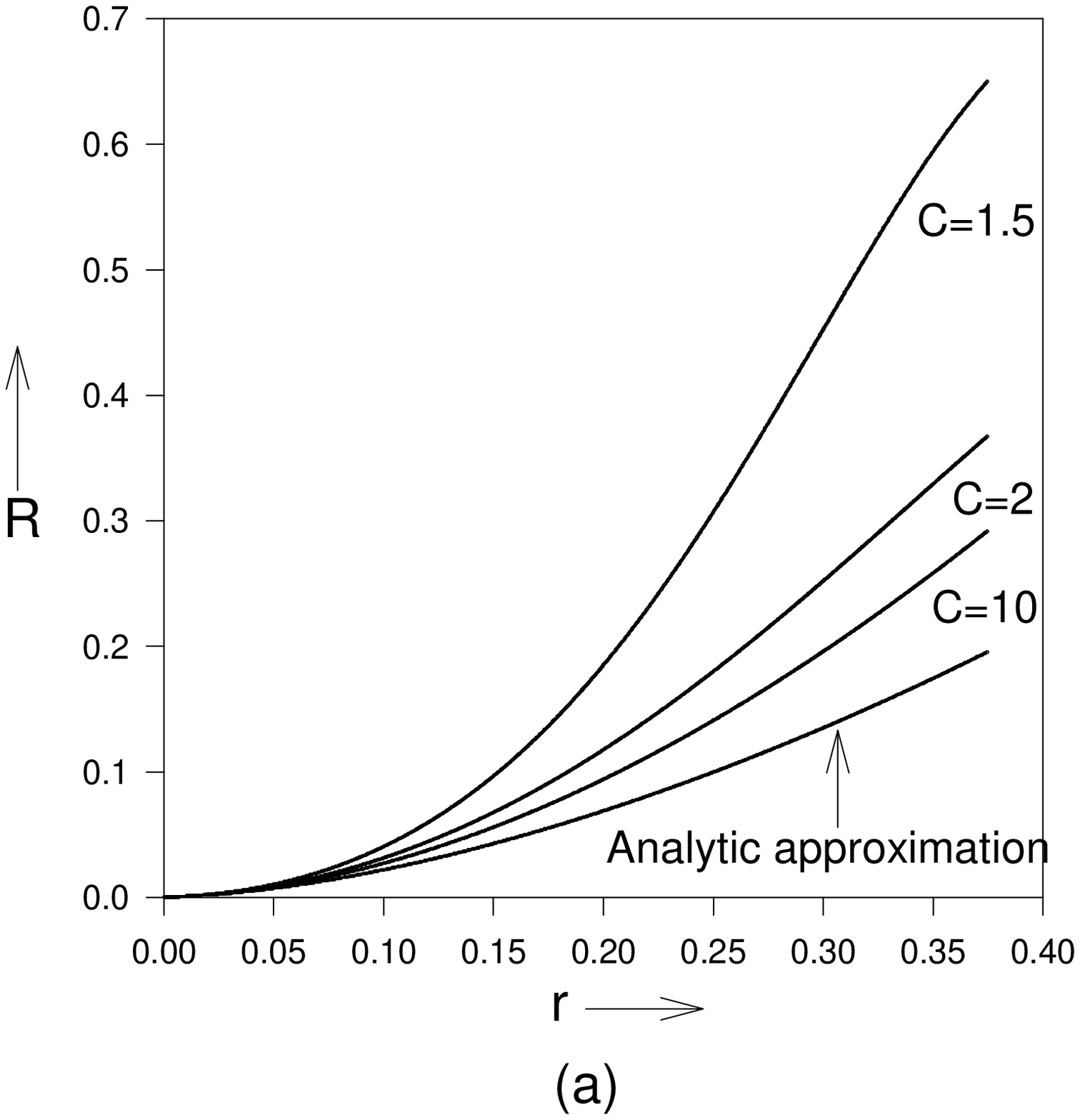}
}
\  \  \
\parbox[b]{7.99cm}
{
\epsfxsize=7.95cm
\epsfbox{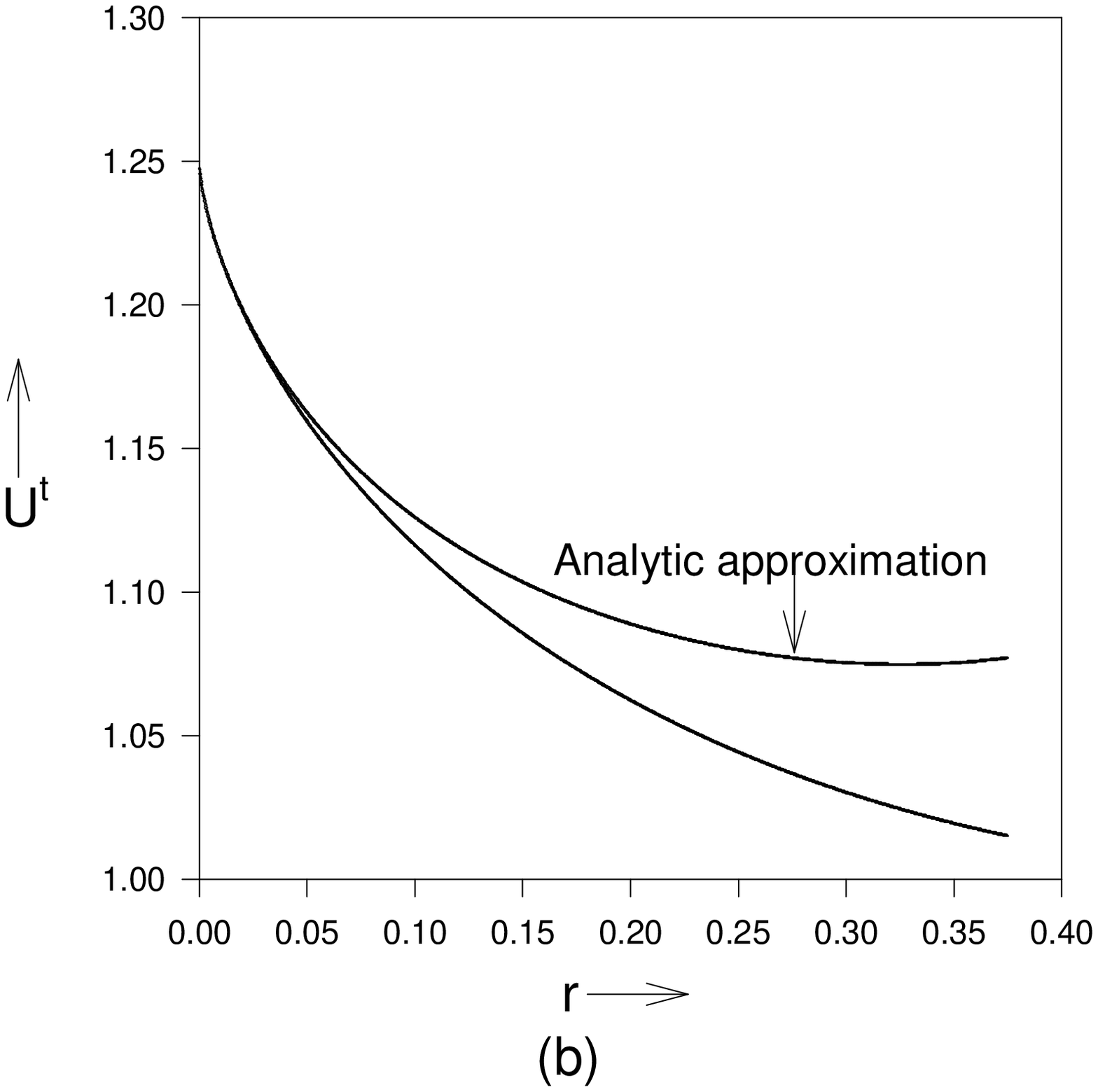}
}
\caption{
Typical ingoing radial timelike geodesics terminating at the central (r=0) 
singularity in the TBL model. The analytic approximation is also shown. 
Here $f=0$ and $\rho= 3-4.8r$.  (a) Different geodesics in 
the (r,R) plane. (b) $U^t(r)$ for C=2 geodesic.
}
\end{figure}   

The scenario which results from the above considerations is that 
a single ingoing timelike geodesic given by Eq. (12), and infinitely 
many other radial nonspacelike geodesics always terminate in future 
at the central singularity. This is regardless of the form of the initial 
data, which in the dust collapse is in the form of the initial distribution 
of the density and velocities at the onset of collapse. 
To understand better the structure of families (outgoing or ingoing) at
the singularity, we note that in case $\rho'(0)\ne 0$ or $\rho''(0)\ne 0$ 
there are always families of ingoing and outgoing radial timelike geodesic 
curves that terminate at the naked central singularity. In cases where 
$\rho'(0)=\rho''(0)=0,\rho'''(0)\ne 0$ there is a critical value of $\rho'''$ 
which determines if there would be outgoing null geodesic families; however, 
the ingoing family would still be there. Thus, in all cases (naked or 
covered), there is a nonzero measure of families of both timelike and null 
particles which terminate at the singularity.

In order to access the strength of the singularity, consider the 
quantity $\Psi=R_{ab}U^aU^b$ along the central particle following the radial 
timelike geodesic $\gamma(s)$ given by Eq. (12). The energy 
density throughout the spacetime is given by
\be
\epsilon (t,r)={F'\over R^2R'}={\rho(r) t_0(r)^2\over [t_0(r)-t]
                -\{[t_0(r)-t] +(2/3) [rt_0'(r)/t_0(r)]t \} } 
\ee
Using this we get
\be
\Psi=R_{ab}U^aU^b={2\over 3(t_0-t)^2}={2\over 3(s_0-s)^2}.
\ee
Since $\Psi$ diverges as the inverse square of the proper time, it follows
that the timelike geodesic $\gamma(s)$ terminates in a strong curvature 
singularity. This shows, as per the criterion of strength of the singularity
given above, that the central singularity forming in the dust collapse 
is always a 
strong curvature singularity. It follows, by continuity, that the curvature 
growth in the past of the singularity will be equally powerful in the 
limit of approach to the singularity along the other nonspacelike 
trajectories as well which meet the singularity with a tangent closer to 
that of the timelike geodesic $\gamma(s)$ considered above.

A similar calculation as above for the particles following radial
timelike trajectories given by Eq. (12) yields
\be
\lim_{s\to s_0}\left((s-s_0)^2\Psi\right)=\lim_{s\to s_0}\left((s-s_0)^2
R_{ab}U^aU^b\right)
={2\over 3}.
\ee
For the sake of completeness, we note that in the case of non-marginally 
bound dust collapse as well, where $f\ne 0$, the quantity $\Psi$ along the 
timelike geodesics $\gamma(s)$ given by Eqs. (12) and (14) behaves in the 
neighborhood of the singularity as
\be
\Psi=R_{ab}U^aU^b\propto{1\over (s_0-s)^2}.
\ee
As is wellknown, the stability criteria in general relativity have
still not been satisfactorily formulated. However, certain norms for 
stability have come to be well accepted. Amongst them is the stability of 
the occurrence of phenomena against small perturbations 
in the symmetry of the spacetime. Recent work$^{17}$ on the perturbations
of the TBL models shows the stability of the naked singularity in this sense. 
Another important stability criterion for any physical phenomena, 
which is the result of time development of initial physical 
configurations (e.g., the initial densities and velocities), is the
stability against perturbations of these physical parameters. 
This is important from the point of view of observations and measurements, 
and also in the context of cosmic censorship, where the role of initial 
data leading to a naked singularity is quite important.

To examine the stability of naked singularities from such a perspective, 
consider for example the marginally bound collapse scenario, in the case
when for a given (at least $C^2$ differentiable) density profile $\rho(r)$
the singularity is naked. Consider perturbations to this initial 
data of the form ${\rho}\to \rho +\delta \rho$. Since any such generic 
infinitesimal perturbation would require the introduction of  
nonvanishing first or second derivatives of the density [i.e. $\rho'(0)\ne 0$ 
or $\rho''(0)\ne 0$; it is known$^{18}$ that for 
realistic stellar models $\rho''(0)<0$], it follows that the singularity 
always continues to be naked. Furthermore, 
as shown above, in all these cases the strong curvature condition is always 
satisfied regardless of the initial data. It follows that the perturbations 
of initial data do not alter the occurrence of a strong curvature naked 
singularity, which is a stable phenomena in that 
the singularity remains not only naked but strong also. 
Similarly, in nonmarginally bound collapse also, a generic perturbation
of the initial data consisting of the initial density and velocity
profiles will leave the naked singularity to be strong.

We thus see that the naked singularity developing in dust collapse is 
stable against such perturbations. What is seen is that the central 
singularity is always a strong curvature singularity, and infinitely many 
timelike geodesic trajectories terminate at this singularity, whether 
covered or naked. The important 
aspect here is the generality of this feature independent of the initial 
density and velocity profiles. It is of course known from earlier studies 
that this singularity, especially when naked, does exhibit a directional 
behavior for the growth of curvature in the cases such as above, and also 
in other collapse models such as the Vaidya spacetimes, along the radial 
null geodesics families terminating at the naked singularity, in the sense 
that the curvature growth may be different along different families. However, 
even in those cases the naked singularity is always strong as per the 
criterion given by Krolak$^{13}$. Further, in a recent development$^{19}$, 
it has been shown that the redshift of rays from the naked singularity is 
finite or infinite depending upon how the density of the star decreases 
away from the center. Since the considerations here show the nature
of the singularity to be stable and strong, the central naked singularity 
formed in collapse becomes a good candidate for a physical phenomena 
likely to have an observational signature. Though we have only considered  
dust here, these results seem to hold for more general equations of state 
as well. This will be discussed elsewhere.

\end{document}